\documentclass[MSNbibl,nameyear,dvips]{arxstspdf}
\usepackage{alltt}
\usepackage{flushend}
\usepackage{stfloats}
\usepackage{graphicx,url,breakurl}
\volume{29}
\issue{2}
\pubyear{2014}
\firstpage{167}
\lastpage{180}
\doi{10.1214/13-STS452} 

\makeatletter

\def\Language#1{{\textsf{#1}}}
\def\LanguageI#1{\LanguageIndex{#1}\Language{#1}}
\def\LanguageIndex#1{\index{#1@\Language{#1}}}
\def\SFunction#1{{\SExpression{#1()}}}
\def\SOperator#1{``{\SExpression{#1}}''}
\def\SClass#1{``{\SExpression{#1}}''}
\def\SPackage#1{\Language{#1}}
\def\SExpression#1{\code{#1}}
\def\SFile#1{\Cq{#1}} 

\def\Cq#1{``\code{#1}''}

\def\Cpp{\LanguageI{C}{$++$}}
\def\C{\LanguageI{C}}
\def\Java{\LanguageI{Java}}
\def\Fortran{\LanguageI{Fortran}}
\def\Lisp{\LanguageI{Lisp}}
\def\R{\Language{R}}
\def\S{\LanguageI{S}}
\def\Python{\LanguageI{Python}}
\def\dQuote#1{``#1''}
\def\Unix{\LanguageI{UNIX}}
\def\Smalltalk{\LanguageI{Smalltalk}}
\def\Simula{\LanguageI{Simula}}

\newcommand{\Twiddle}{\mbox{\(\mathtt\sim\)}}

\def\code#1{\texttt{#1}}

\def\chCM#1{\textit{#1}}

\makeatother

\begin{document}
\begin{frontmatter}

\title{Object-Oriented Programming, Functional Programming and \R{}}
\runtitle{Object-Oriented and Functional Programming}

\begin{aug}
\author{\fnms{John M.} \snm{Chambers}\corref{}\ead[label=e1]{jmc@stat.stanford.edu}}
\address{John M. Chambers is Consulting Professor, Department of
Statistics, Stanford University, Stanford, California 94305-4065, USA
\printead{e1}.}
\affiliation{Stanford University}
\runauthor{J. M. Chambers}

\end{aug}

%
\begin{abstract}
This paper reviews some programming techniques in \R{} that have
proved useful, particularly for substantial projects. These include
several versions of object-oriented programming, used in a large
number of \R{} packages. The review tries to clarify the origins
and ideas behind the various versions, each of which is valuable in
the appropriate context.

\R{} has also been strongly influenced by the ideas of functional
programming and, in particular, by the desire to combine functional
with object oriented programming.

To clarify how this particular mix of ideas has turned out in the
current \R{} language and supporting software, the paper will first
review the basic ideas behind object-oriented and functional
programming, and then examine the evolution of \R{} with these ideas
providing context.

Functional programming supports well-defined, defensible
software giving reproducible results.
Object-oriented programming is the mechanism \emph{par excellence} for
managing complexity while keeping things simple for the user.
The two paradigms have been valuable
in supporting major software for fitting models to data and numerous
other statistical applications.

The paradigms have been adopted, and adapted, distinctively
in \R{}. Functional programming motivates much of \R{} but \R{}
does not enforce the paradigm.
Object-oriented programming from a functional perspective differs
from that used in non-functional languages,
a distinction that needs to be emphasized to avoid confusion.

\R{} initially replicated the \S{} language from Bell Labs, which in turn
was strongly influenced by earlier program libraries. At each
stage, new ideas have been added, but the previous software
continues to show its influence in the design as well.
Outlining the evolution will further clarify why we currently have
this somewhat unusual combination of ideas.\looseness=-1
\end{abstract}

%
\begin{keyword}
\kwd{Programming languages}
\kwd{functional programming}
\kwd{object-oriented programming}
\end{keyword}

\end{frontmatter}

\section{Introduction}
\label{sec:introduction}

\R{} has become an important medium for communicating new methodology
in statistics and related technology. References to the supporting
\R{} software
frequently accompany journal articles or other publications
describing new results.
The software is available to other \R{} users,
ideally as a package in a standard repository.
The benefits for statistics as a discipline are considerable: The
community has rapid access to new ideas in a free, open-source format
as software that can in most cases be installed and used immediately by
those interested in the statistical techniques.
The user community has both created and benefited from this resource.

This paper examines two of the most significant paradigms
in programming languages generally: object-oriented programming (OOP)
and functional programming.
\R{} makes use of both, but in its own way.
Both paradigms are valuable for serious programming with the language.
But in both cases, understanding the relevant ideas in the context of
\R{} is needed to avoid confusion.
The confusion sometimes arises, in both cases, from applying to
\R{} interpretations of the paradigms that apply to other languages
but not to this one.
Section~\ref{sec:main-ideas} of the paper will review the ideas,
generally and in their
\R{} versions, with the goal of clarifying the basics.
Given the importance of \R{} software to the community, creators of
new \R{} software should benefit from
understanding these concepts.

We will also examine in Section~\ref{sec:evolution} of the paper the
evolution that led to these versions of
functional programming and OOP.
The prime motivation was not language design in the abstract but
to provide the tools needed for
research and data analysis by the user community at the time.
\R{} originally reproduced the functionality of the \S{}
language at Bell Labs, which itself had evolved through several stages
beginning in the late 1970s
and which was in turn based on earlier statistical software libraries,
mainly in \Fortran{}.

\R{} added important new ideas and has continued to evolve, but the
main contents inherited through \S{} shaped the capabilities and the
approach to statistical computing.
In a surprising number of areas, what we think of as \dQuote{the \R{}
way} of organizing the computations actually reflects software
developed twenty years or more before
\R{} existed.

Having been involved in all the stages, I am naturally inclined to
a historical perspective,
but it is also the case that the history itself had
substantial impact on the results.
It may be comforting to view programming languages as abstract
definitions, but in practice they evolve from the needs, interests and
limitations of their creators and users.


\section{Functional and Object-Oriented Programming:
The Main Ideas}
\label{sec:main-ideas}

Functional and
object-oriented programming fit naturally into statistical
applications and into \R{}.
The original motivating use case, fitting models to data, remains
compelling.
An expression such as\vspace*{6pt}
\begin{list}{}{\setlength{\leftmargin}{1em}}
\item{}
\begin{alltt}\small
irisFit <- lm(Sepal.Width \Twiddle{}
              . - Sepal.Length, iris)
\end{alltt}
\end{list}\vspace*{6pt}
calls a function that creates an object representing the linear model
specified by the first argument, applied to the data specified by the
second argument.
The computation is functional, well-defined by the arguments.
It returns an object whose properties provide the information needed
to study and work with the fitted model.
Other functions and other objects can adapt to different models in a
form that is convenient for both the user and the implementer.

Principles of functional programming guide us in writing reliable,
reproducible functions for the different models.
Object-oriented programming provides tools for defining the model
objects clearly, and adapting to new ideas and new forms of models.
Section~\ref{sec:from-functions-oop} goes into details of the
\R{} implementations.

As they have been realized in \R{}, both paradigms center on a few,
intuitive concepts.
The details are more complicated, as they usually are.
In the case of functional programming, the realization in \R{} is only
partial, reflecting the language's origins as well as practical
considerations.
In the case of OOP, there are now at least three realizations of the
ideas in \R{}, using two
different paradigms.
All three have significant applications and practical value.

Despite all these devilish details, the main ideas remain visible and
useful, particularly when programming serious applications using the language.

\subsection{Functional Programming}
\label{sec:main-funct-progr}

For our purposes, the main principles of functional programming can be
summarized as follows:
\begin{longlist}[1.]
\item[1.] Programming consists largely of defining \emph{functions}.
\item[2.] A function definition in the language, like a function in
mathematics, implies that a function call returns a unique value
corresponding to each valid set of arguments, but \emph{only} dependent
on these arguments.
\item[3.] A function call has no side effects that could alter other
computations.
\end{longlist}
The implication of the second point is that functions in the
programming language are mappings from the
allowed set of arguments to some range of output values.
In particular, the returned value should not depend on other
quantities that affect the \dQuote{state} of the software when the
function call is evaluated.\looseness=1

True functional languages conform to these ideas both by what they do
provide, such as pattern expressions, and what they do not provide,
such as procedural iteration or dynamic assignments.
The classic tutorial example of the factorial function, for example,
could be expressed in the Haskell language by the pattern:\vspace*{6pt}
\begin{list}{}{\setlength{\leftmargin}{2em}}
\item{}
\begin{alltt}\small
factorial x = if x > 0
  then x * factorial (x-1) else 1,
\end{alltt}
\end{list}\vspace*{6pt}
plus some type information, such as that a value for \SExpression{x}
must be an
integer scalar.

Is \R{} a functional programming language in this sense? No.
The structure of the language does not enforce functionality;
Section~\ref{sec:in-R} examines that structure as it relates to
functional programming and OOP.
The evolution of \R{} from earlier work in statistical computing also
inevitably left portions of earlier pre-functional computations;
Section~\ref{sec:evolution} outlines the history.
Random number generation, for example, is implemented in a distinctly
\dQuote{state-based} model in which an object in the global
environment (\SExpression{.Random.seed}) represents the current state
of the generators.
Purely functional languages have developed techniques for many of
these computations, but rewriting \R{} to eliminate its huge body of
supporting software is not a practical prospect and would require
replacing some very well-tested and well-analyzed computations (random
number generation being a good example).

Functional programming remains an important paradigm for statistical
computing in spite of these limitations.
Statistical models for data, the motivating example for many features
in \S{} and \R{}, illustrate the value of analyzing the software from
a functional programming perspective.
Software for fitting models to data remains one of the most active
uses of \R{}.
The functional validity of such software is important both for
theoretical justification and to defend the results in areas of
controversy: Can we show that the fitted models are well-defined
functions of the data, perhaps with other inputs to the model such
as prior distributions considered as additional arguments?
The structure of \R{} as described in Section~\ref{sec:in-R} can
provide support for analyzing functional validity.
Equally usefully, such analysis can also illuminate the limits of
functional validity for particular software, such as that for model-fitting.

\subsection{Object-Oriented Programming}
\label{sec:main-OOP}

The main ideas of object-oriented programming are also quite simple and
intuitive:
\begin{longlist}[1.]
\item[1.] Everything we compute with is an \emph{object}, and objects
should be
structured to suit the goals of our computations.
\item[2.] For this, the key programming tool is a \emph{class} definition saying
that objects belonging to this class share structure defined by
\emph{properties} they all have, with the properties being themselves
objects of some specified class.
\item[3.] A class can \emph{inherit} from (contain) a simpler superclass, such
that an object of this class is also an object of the superclass.
\item[4.] In order to compute with objects, we can define \emph{methods}
that are only used when objects are of certain classes.
\end{longlist}
Many programming languages reflect these ideas, either from
their inception or by adding some or all of the ideas to an existing
language.

Is \R{} an OOP language? Not from its inception, but it has added
important software reflecting the ideas.
In fact, it has done so in at least three separate forms, giving rise
to some confusion that this paper attempts to reduce.

Some of the confusion arises from not recognizing that the final item
in the list above can be implemented in radically different ways,
depending on the general paradigm of the programming language.
A key distinction is whether the methods are to be embedded in some
form of
functional programming.

Traditionally, most languages adopting the OOP paradigm are not
functional; either the language began with objects and classes as a
central motivation (\Language{SIMULA}, \Java{}) or added the paradigm
to an existing non-functional language (\Cpp{}, \Python{}).
In such languages, methods were naturally associated with classes,
essentially as callable properties of the objects.
The language would then include syntax to call or \emph{invoke} a
method on a particular object, most often using the infix operator \Cq{.}.
The class definition then encapsulates all the software for the class.
Where methods are needed for other computations, such as special
method names in \Python{} or operator overloading in \Cpp{},
these are provided by ad-hoc mechanisms in the language, but the
method remains part of the class definition.

In a language that is functional or that aspires to behave
functionally as \S{} and \R{} do, the natural role of methods
corresponds to the intuitive meaning of \dQuote{method}---a technique
for computing the desired result of a function call.
In functional OOP, the particular computational technique is chosen
because one or more arguments are objects from recognized classes.

Methods in this situation belong to functions, not to classes; the
functions are \emph{generic}.
In the simplest and most common case, referred to as a standard generic
function in \R{}, the function defines the formal arguments
but otherwise consists of nothing but a table of the
corresponding methods plus a command to select the method in the table
that matches the classes of the arguments.
The selected method is a function; the call to the generic is then
evaluated as a call to the selected method.

We will refer to this form of object-oriented programming as
\emph{functional} OOP as opposed to the \emph{encapsulated} form in
which methods are part of the class definition.

\subsection{Their Relationship to \R{}}
\label{sec:in-R}

To understand computations in \R{}, two slogans are helpful:
\begin{itemize}
\item Everything that exists is an object.
\item Everything that happens is a function call.
\end{itemize}
In contrast to languages such as \Java{} and \Cpp{} where objects are
distinct from more primitive data types, every reference in \R{} is to
an object, in particular, to a single internal structure type in the underlying
\C{} implementation.
This applies to data in the usual sense and also to all parts of the
language itself, such as function definitions and function calls.
Computations that are more complex than a constant or a simple name
are all treated as function calls by the \R{} evaluator, with control
structures and
operators simply alternative syntax hiding the function call.
[Details and examples are shown in (\cite{R:Chambers:2008}, pages 458--468).]

The two slogans, however, do not imply that computations in \R{}
must follow either
functional or object-oriented programming in the senses outlined in
the preceding sections.
With respect to object-oriented programming, \R{} has several
implementations that have evolved as outlined in Section~\ref{sec:evolution}.
These can be used by programmers to provide software following either
of the OOP paradigms.

Functional programming's relationship to \R{} is less straightforward.
The evaluation process in \R{} does not enforce functional
programming, but does encourage it to a degree.
In particular, the evaluation process in \R{} contributes to
functional programming by largely avoiding side effects when function
calls are evaluated, but some mechanisms in the language and
especially in the underlying support code can behave in a non-functional way.
To understand in a bit more detail, we
need to examine this evaluation process.

Computations in \R{} are carried out by the \R{} evaluator
by
evaluating function call objects.
These have an expression for the function definition (usually a
reference to it by name) and zero or more expressions for the
arguments to the call.
The full details are somewhat beyond our scope here, but an essential
question is how references to objects are handled.
Any programming language must have references to data, which in \R{}
means references to objects.
As discussed in Section~\ref{sec:evolution}, the evolution of such
references is central to the evolution of programming languages,
especially for statistics.

In \R{} a reference to an object is the combination of a name and a
context in which to look up that name; the contexts in \R{} are
themselves objects, of type \Cq{environment}.
A reference is therefore the combination of a name and an environment.
(We'll look at an example shortly.)

Note that we are talking about references \emph{to} objects; most
objects in \R{} are not themselves reference objects.
Languages implementing OOP in the traditional, non-functional form
essentially always include reference objects, in particular, what are
termed \emph{mutable} references.
If a method alters an object, say, by assigning new values to some of
its properties, all references to that object see the change,
regardless of the context of the call to the method.
Whether the reassignment of the property takes place where the object
originated or down in some other method makes no difference; the
object itself is the reference.

In contrast, the reference in \R{} consists of a name and an
environment---the environment in
which the object referred to has been assigned with that name.
Most \R{} programming is based on a concept of \emph{local
references}; that is, reassigning part of an object referred to by
name alters the object referred to by that name, but only in the local
environment.
If that local reference started out as a reference in some other
environment, that other reference is still to the original object.

To understand the relation of local references to functional
programming in
\R{}, an example and a few more details of function call evaluation
are needed.
\R{} evaluates function calls as objects.
For example, when the evaluator encounters the call\vspace*{6pt}
\begin{list}{}{\setlength{\leftmargin}{2em}}
\item{}
\begin{alltt}\small
lm(Sepal.Width
   \Twiddle{} . - Sepal.Length, iris),
\end{alltt}
\end{list}\vspace*{6pt}
it uses the object representing the call to
create an environment for the evaluation.

The call identifies the function, also an object of course, typically
referring to it by name.
In this case \SExpression{lm} refers to an object in the
\SPackage{stats} package.
That object has formal arguments [14 of them, in the case of
\SFunction{lm}].
The evaluator initializes an environment for the call with objects corresponding
to the formal arguments, as unevaluated expressions built from the two
actual arguments and default expressions found in the function
definition.
For details see Section~4 of the language definition,
\citet{RLang} and Chapter~13 of \citet{R:Chambers:2008}.
As an aside, the common use of terms like \dQuote{call by value}
(and the contrasting \dQuote{call by reference}) for argument passing
in \R{} is invalid and
misleading.
Arguments are not \dQuote{passed} in the usual sense.

Local references operate on all the objects in the environment to
prevent side effects.
The formal argument \SExpression{data} to \SFunction{lm} matches the
expression \SExpression{iris}, which refers to an object in the
\SPackage{datasets} package.
Expressions that extract information from \SExpression{data} work
on that object.
But the local reference defined by \SExpression{data} and the
environment of the evaluation is distinct from the reference to
\SExpression{iris} in the package.
If an assignment or replacement expression is encountered that would alter
\SExpression{data}, the evaluator will duplicate the object first to
ensure locality of the reference.

The local reference paradigm
is helpful in validating the functionality of an \R{} function.
Only the local assignments and replacements need to be examined; calls
to other functions
will not alter references in this environment, so long as those
functions stick to local reference behavior.
If a function \SFunction{f} calls a function \SFunction{g} and both
functions stick to local reference assignments, then knowing that
the value of a call to
\SFunction{g} depends only on the arguments is all that is
needed; how \SFunction{g} computes that value is irrelevant.

While local references help avoid side effects, they do not prevent
computations from referring to objects or other data outside the
functions being called, and therefore potentially returning a result
that depends on a non-functional \dQuote{state.}
Whether a particular computation in \R{} is strictly functional can
only be determined by examining it in detail, including all the
functions that call code in \C{} or \Fortran{}.

The rest of this section takes a slight detour to consider how one
might do that examination.

\subsection*{Validating Functionality in \R{}}

In principle,
the functional validity of particular computations could be analyzed
and either certified or the limitations to functionality reported.
Such functional validation would be useful in cases where either the
theoretical validity or the implications of the result in an
application are being questioned.
Fitting models to data provides a natural example for both aspects.
Given a function taking as arguments data and a
model specification and returning a fitted model object, can one validate
that the returned object is functionally defined by the arguments?
If not, can the non-functionality be parametrized meaningfully, in
which case one can construct a functional version of the computation
by including such parameters as implicit arguments?
\R{} does not have organized support for such validity investigations,
but developing tools for the purpose would be a worthwhile project.

Functional validation is a bottom-up construction.
The bottom layer consists of any functions called that are not
implemented in \R{}, typically those that call routines in \Cpp{},
\C{} or \Fortran{}.
Included are the \R{} primitives, routines from numerical libraries
and a variety of other standard sources, plus any new code brought in
to implement the computation in question.
The functional validity of each of these is an empirical assertion.
Some are clearly non-functional, such as the \mbox{\SOperator{<<-}} operator
and \SFunction{assign} function that do nonlocal assignments.

Many computations in \R{} eventually call subprograms not originally
written for \R{}.
Each of these must be examined for potential non-functional behavior,
sometimes a daunting task.
However, good practice in using well-tested, preferably open-source
supporting software will often provide a plausible basis.

If \R{} code includes an interface to code in \C{},
\Fortran{} or other languages whose functional validity cannot be
established, nothing more can be said.
Other than such code, functional
validity is likely to fail for one of three reasons:
\begin{itemize}
\item dependance on nonlocal values;
\item using low-level computations in \R{} known to violate functionality;
\item changing functions or other objects at run time.
\end{itemize}
A prime example of the first is the use of external data, such as the
global options object, for convergence tolerances or other
parameters for iterative numerical computations.
An example of the second is the inclusion of pseudo-random values in
the calculation.
The third problem might be caused, for example, by using a function
from the global environment.

The third danger is greatly reduced when the code resides in the
namespace of a package
with explicit import rules.
Any reasonable approach to validating functionality would make this a
requirement.

My feeling is that most examples of failures could be corrected to
create functionally
valid extensions of the computation in question.
Tolerances are often organized through the \R{} \SFunction{options}
function, explicitly designed to avoid functional programming by
allowing users to set state parameters that are then queried by the
calculation.
Once identified, such options could be converted to additional
arguments to the function being validated.
[A general mechanism would be a version of \SFunction{getOption} that
required the option in question to be supplied as an argument.]

Pseudo-random values are used in a variety of procedures, including
some optimization techniques where they are expected to provide more
robust numerical behavior by jittering values during iteration.
These can be made functionally valid by using well-defined generator
software, such as that supplied in \R{} itself, and by treating the
initial state of the generator as another nonlocal value to be
incorporated as an additional argument.
One should always include an explicit initialization via
\SFunction{set.seed} in any example expected to be reproducible, and
that practice can be the basis for a functionally valid version of the
computation.

Beyond these specific examples, numerical computations often
depend on the underlying parameters of the floating-point
computations, for example, to select convergence criteria for
iteration. Fortunately, several decades of work by numerical analysts
and hardware designers have greatly standardized
the specification of
the numerical engine in modern computers: just knowing 32-bit or 64-bit
gets us a long way.

Developing a framework for validating functionality seems to me an
interesting cooperative research direction that could be of value to
the statistical community.


\section{The Evolution of Functional Programming, OOP and \R{}}
\label{sec:evolution}

The computational paradigms for functional programming and for
object-oriented programming have evolved from a sequence of changes in
software, beginning with the earliest programable computers.
During the same period, software for statistics was also evolving, one
thread of which led through early libraries to \S{} and then to \R{}.

There may be an appearance of earlier languages being replaced by
later and presumably improved approaches.
It is true that each major revision asserts improvements that will
extend our abilities to express our ideas in software.
However, none of the versions of \S{} or \R{} actually totally
replaced earlier software paradigms.

The current software in, and interfaced from, \R{} illustrates this
evolution.
\R{} has developed important new techniques, but originated from
the \S{} language, reproducing nearly all of \S{} as it was
described at that time.
\S{} in turn went through several evolutionary changes and was itself
based on extensive earlier software, particularly subroutine libraries
for \Fortran{} programming.
Examining the history shows that a surprising portion of what we see
now is structure inherited from the early stages.

The form in which functional programming and OOP were adopted was also
influenced by the existing software.
Examining the history will explain many of the choices made.


\subsection{From Hardware to Data and Libraries}
\label{sec:from-storage-objects}

\begin{figure*}

\includegraphics{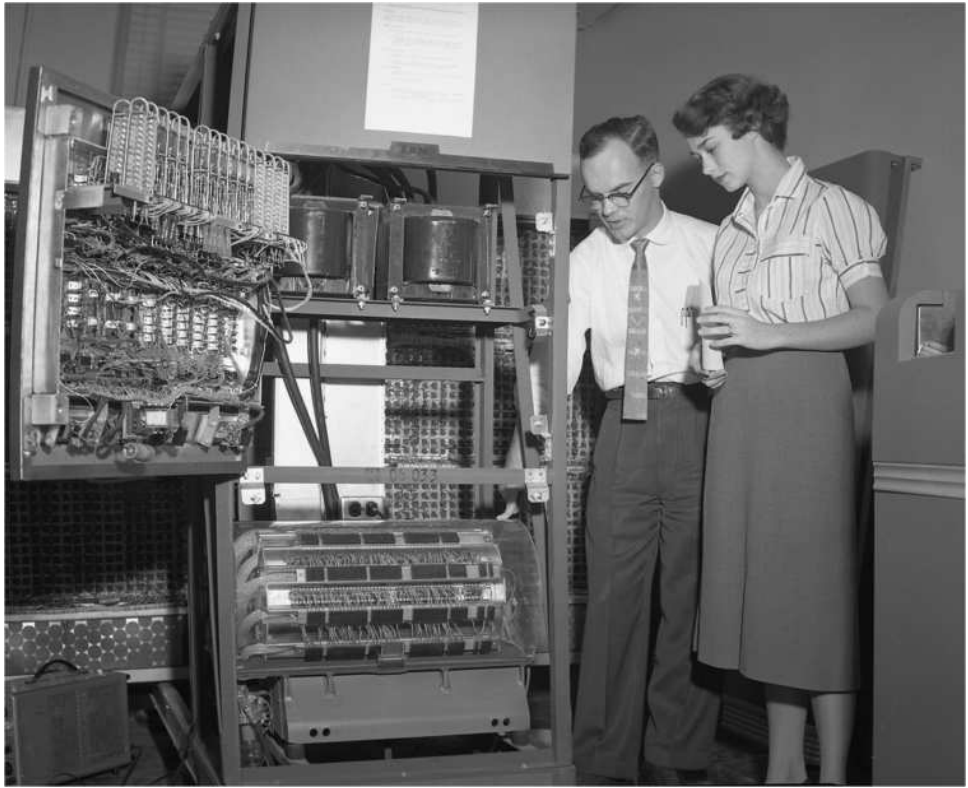}

\caption{An IBM 650 computer, mid 1950s. Under the glass is the
magnetic drum storage unit (memory), 2000 words for data and programs.}
\label{fig:650}\vspace*{-6pt}
\end{figure*}

The earliest general-purpose computers were programmed in terms of the
physical machine, its storage and the basic operations provided to
move data around and perform arithmetic and other operations.
The IBM 650
(Figure~\ref{fig:650})
was probably the first computer widely sold and used (and
the machine on which I did my first programming, around 1960).

In this pre-silicon world, storage for data or programs resided on a
rotating magnetic drum, holding 2000 decimal words.
Data could be read or written only when the corresponding segment of
the drum passed under the appropriate fixed head, so that physical
positioning of data was a serious aspect of performance.
With this close view of the hardware, programming languages (assembly
languages for the actual machine instructions) defined storage in
terms of single physical units (words in the 650) and blocks of
sequential storage.

This was not an environment to encourage abstraction of ideas about
data.
However, by 1960 the first generation of \dQuote{high-level} languages
had been
introduced and would support profound changes.
For statistical computing this meant primarily \Fortran{}.

In terms of data storage, \Fortran{} actually continued the basic
notion of single items (scalars) and contiguous blocks (arrays).
Two major changes, however, were made.
First, the contents were described in terms of their content, the
first \emph{data types} including integer and floating point numbers.
Second, the language encouraged operations that iterated over the
contents of the arrays.
By interpreting an array as a sequence of equal-length subarrays, this
indexing extended to matrices and to multi-way tables.

Along with the new paradigm for data and facilities for iteration, the
high-level languages
encouraged software to be organized in subroutines, so that a
computational method could be realized as one or several units of
software.
While the changes may seem modest from the current perspective, they
in fact supported a major revolution in scientific computing generally
and emphatically so in computing for statistics.

Algorithm series and other publications supported by professional
societies began
to accumulate refereed, trustworthy procedures for many key
computations.
The statistics research group at Bell Labs developed a large
\Fortran{} library that reflected our needs and our philosophy of
research and data analysis.
The book \dQuote{Computational Methods for Data Analysis},
\citet{CompMethods}, did not present software but did reflect
the tools that would later form the basis for \S{}.
After an introduction and discussion of program design, the remaining
six chapters covered computations supported by the library:
\begin{longlist}[3.]
\item[3.]\chCM{Data Management and Manipulation}
(including sorting and table lookup).
\item[4.]\chCM{Numerical Computations}
(approximations,\break  Fourier transforms, integration).
\item[5.]\chCM{Linear Models}
(numerical linear algebra, regression, multivariate methods).
\item[6.]\chCM{Nonlinear Models}
(optimization, nonlinear least squares).
\item[7.]\chCM{Simulation of Random Processes}
(random number generation and Monte Carlo).
\item[8.]\chCM{Computational Graphics}
(plotting techniques, scatter plots, histograms and probability plots).
\end{longlist}
Each of these was supported in the pre-\S{} era by subroutines that
would then become the basis for corresponding functions in \S{}.

Much of the organization for basic tools in \R{} has inherited, through
\S{}, the structure of the subroutine library.
That includes the graphical computations, in particular, features
essential to \S{} and \R{}:
separation of graphic device specification from plotting; the plot, figure
and margins structure; graphical parameter specification to control
style.
These were not created for \S{} but taken over from previous
\Fortran{} software, described in
\citet{gr-z}.

The Bell Labs software was in the background of
\citet{CompMethods}, but general readers were given instructions for
obtaining similar software from publicly available sources
for the methods described.
The procedure would not always be simple, but the potential
availability marked
a big step forward.
For the first time, statisticians could draw on an extensive range of relevant
software to support their research, at least in principle.
Various statistical software packages had existed for some time,
but these were by and large oriented to routine analysis, to teaching
or to specialized statistical techniques.
\citet{CompMethods} and the software it reflected were aimed at research
in statistics
and challenging data analysis.
For this purpose, a more general and open-ended approach was needed.

\subsection{From \Fortran{} to \S{}}
\label{sec:from-fortran-s}

For those involved with statistical theory or applications, in academia or
industry,\vadjust{\goodbreak} there were two main limitations to the software described so
far: availability and the
programming interface.
The Appendix to \citet{CompMethods} was a set of tables for
each of the chapters, with rows corresponding to computational tools
that were more or less available to readers.
The last column of the table listed sources for the corresponding software.
The entries in that column were not uniformly helpful; in the best
situation, a generally
available program library could be ordered that provided a number of
the subroutines, but these were not designed for statistical
applications, most being directed at numerical methods typically
motivated by applications in physics.
More than half of the entries
read \dQuote{Listing,} implying a laborious and error-prone manual
procedure for the user.
[As an example, many \dQuote{bug reports} came to us as a result of
confusing an \Cq{I} and a \Cq{1} when typing in the stable
distribution software,~\citet{stable}.]\looseness=1

Substantial in-house libraries, such as the one at Bell Labs,
gave users a fairly wide range of computations, supported by
improved numerical and other algorithms.
However, to apply the computations specifically to a particular
dataset with particular results in mind required some substantial
additional \Fortran{} programming.
That programming had to be repeated and revised for each analysis or
research question.

In the 1970s the situation was therefore a combination of improved
basic computational capabilities but with a high programming barrier for
most statisticians.
The classical linear regression in \Fortran{} as shown in \citet
{SBook85}, for example, was fairly straightforward:\vspace*{6pt}
\begin{list}{}{\setlength{\leftmargin}{2em}}
\item{}
\begin{alltt}\small
call lsfit(X, N, P, y, coef, resid).
\end{alltt}
\end{list}\vspace*{6pt}
This computes the fitted model and returns it as vectors of
coefficients and residuals.
The data as objects are restricted to arrays, a matrix \SExpression{X}
and vector \SExpression{y} for the data and two arrays,
\SExpression{coef} and \SExpression{resid} for the fitted model.
The structure of the objects and their storage allocation remains the
programmer's responsibility.
Linking the basic computation to the data in an actual analysis
remained nontrivial and mistakes along the way were likely.
And this is for the most standard of models.
Even given an extensive library, the programming to apply the tools to
most applications was a laborious, error-prone activity, usually assigned
to dedicated programmers, research assistants or students.
The statistician's ideas went through nontrivial translation before they
were expressed as computations.

The first two versions of \S{} were designed to provide an
\dQuote{interactive environment} that included the computational areas
described
in \citet{CompMethods} and that allowed the statistician to formulate
ideas directly for computation.
The second version of \S{} was licensed for general use and described in
\citet{SBook84}.

In \S{}, the linear regression computation became a simpler
expression, storage
for data was provided automatically and the
returned model was now an object, with components for the coefficients
and residuals:\vspace*{6pt}
\begin{list}{}{\setlength{\leftmargin}{2em}}
\item{}
\begin{alltt}\small\label{page:functionalReg}
fit <- reg(X, y).
\end{alltt}
\end{list}
At this stage, \S{} had a functional appearance, not radically unlike
\R{},
but its paradigm was essentially an extension of the \Fortran{} view.
Dynamically created, self-describing
objects were assigned in a single workspace, but
the underlying computations were those of the earlier subroutine
library: The functions in \S{}, documented in \citet{SBook84}, were in
fact interfaces to \Fortran{} subroutines: \SFunction{reg} would in
fact be programmed by calling \SFunction{lsfit}.

Although there was a macro facility in the language, programming a
function in this version of \S{} meant \dQuote{extending \S{}} as
described in the book of that name, \citet{SBook85}.
The definition of the new function was programmed in an
\dQuote{interface language} built on \Fortran{} and compiled from its
\Fortran{} translation.
As the main programming mechanism this was unsatisfactory, in the
sense that extending the language had a substantial learning barrier
beyond using the language.
The
ability to access other software via an inter-system interface remains
a key feature of \R{}, however, one still under active development.

Equally as important as the technical side was the beginning of a
network of statisticians involved in creating and sharing software
through the medium of the language.
\S{} was licensed from the early 1980s, available thanks to the newly
distributed \Unix{} operating system,
with inexpensive academic
licenses to encourage adoption by university researchers, also following
the example of \Unix{}.
Open-source software was not an option, but the research community was
increasingly involved and their interest stimulated further
developments on our part, particularly from contacts with interested
users belonging to a \dQuote{beta
testing} network.

Simultaneously, we were thinking about a new approach to the language
itself, emphasizing
the programming aspect of creating new software for statistical and
other quantitative applications.
Described initially in \citet{QPE} as a language separate from~\S{},
this research later merged with
other changes to form the next version, labeled S3 and
described in the \dQuote{blue book,} \citet{SBook88}.
The slogans in Section~\ref{sec:in-R} were basic to this version of \S
{}: everything is an
object (stated explicitly) and function calls do all the computation
(implicit).

This was functional programming (more or less) and object-\emph{based}
but not object-oriented.
Objects were given structure through attributes attached to
vectors and through named
components, but there were no classes or methods.

\subsection{From Data to Classes and Methods}
\label{sec:from-data-classes}

The languages that originated the concepts of classes, properties,
inheritance and methods came out of several motivations.
The first, \Simula{}, was concerned with simulating systems.
In retrospect, modeling by simulation and modeling by fitting to data
have clear
correspondences but with quite a different perspective.
For an example, suppose we want to simulate a simple model for an
evolving population of individuals.
In \R{} notation, but quite in the style of \Simula{}, we define a
class \SExpression{SimplePop}.
An object from this class is a specific realization of the model
population with properties that define the probabilities of birth and
death, and a vector of population size at each generation.
An object from the population is created by calling the generator for
the class:\vspace*{6pt}
\begin{list}{}{\setlength{\leftmargin}{2em}}
\item{}
\begin{alltt}\small
p <- SimplePop(birth = 0.08,
               death = 0.1,
                  size = 100).
\end{alltt}
\end{list}\vspace*{6pt}
Rather than a single functional computation as in the case of linear
regression, computations proceed by simulating the evolution of the
population object \SExpression{p}.
The object itself evolves; in the terminology of OOP, it is a \emph{mutable
reference}.

A corresponding difference in the programming paradigms of \S{} and the
emerging OOP languages was that the latter did not take a functional
view of computation.
Instead, computations largely consisted of
invoking a method on an object.
In the \SExpression{SimplePop} example, the fundamental computation is to
simulate one generation of the evolution by invoking the \SFunction
{evolve} method\vspace*{6pt}
\begin{list}{}{\setlength{\leftmargin}{2em}}
\item{}
\begin{alltt}\small
p$evolve().
\end{alltt}
\end{list}\vspace*{6pt}
%
The value returned by this method is irrelevant.
The method's purpose is to change the object, in this case by
simulating one further generation and appending the resulting value to
a property in the object, namely, \SExpression{p\$size}. (See files
\SFile{SimplePop.R} and
\SFile{SimplePopExample.R} in the supplementary materials.)

Following the development of \Simula{} in the late 1960s, a variety of
languages adopted this paradigm.
\Cpp{} added classes and methods to the \C{} language; like \C{}, it
was initially used for a
variety of programming tasks implementing \Unix{} and application
software for \Unix{}.
In contrast to the \dQuote{add-on} nature of \Cpp{}, the \Smalltalk{}
language was a very pure, simplified realization of the ideas in
\Simula{}.
Its major, and revolutionary, application was to implement the
graphical user interface created at Xerox PARC in the 1970s.
Many other versions of encapsulated OOP followed, either
added on to existing languages or incorporated into new languages from
the start.

Dialects of the \Lisp{} language and languages based on \Lisp{} also
incorporated OOP in various forms.
During the 1980s, several research projects built statistical software
on the basis of these languages, including
some elegant and potentially widely
applicable systems, notably LISP-STAT, \citet{LispStat}.
As it turned out, however, the most widely used version of OOP for
statistical applications would come from a somewhat casual approach in
\S{}.

\subsection{Functional OOP in \S{} and \R{}}
\label{sec:from-functions-oop}

The chief motivation for introducing classes and functional methods to
\S{} was the initial application: fitting, examining and modifying diverse
kinds of statistical models for data.
This remains arguably the most compelling example for functional OOP
in statistics.
The \dQuote{Statistical Models in S} project reported in \citet
{SMS}---the \dQuote{white book}---brought together ten authors
presenting software for a variety of
statistical models, from linear regression to tree-based models.
The different models were presented as consistently as possible.

Each type of model had a definition as an object having the information,
such as coefficients and other properties, required.
The object was created by a corresponding function taking as arguments
the data, model description
and possibly other controlling parameters.
A linear regression fit, for example, called the function \SFunction{lm}:\vspace*{6pt}
\begin{list}{}{\setlength{\leftmargin}{0em}}
\item{}
\begin{alltt}\small
irisFit <- lm(Sepal.Width
              \Twiddle{} . - Sepal.Length, iris)
\end{alltt}
\end{list}\vspace*{6pt}
and returned a corresponding linear regression object.
Further computations on this object would examine the model, return
information about it, or update the fit.
The underlying computations still used basic software similar to that
for \SFunction{lsfit} and \SFunction{reg}.
However, the description of the model (a formula) and the data (a~data
frame) were designed to apply to statistical models generally.
For example, to fit a generalized
linear model the user called \SFunction{glm} with formula and data
arguments typically similar to those in a call to \SFunction{lm}.
Other arguments would provide information suitable to the particular
type of model (a \SExpression{link} function, e.g.).

For the convenience of the user, further computations should have a
uniform appearance.
To print or plot the fitted model or to compute predictions or an
updated model corresponding to new data, the user should call the same
function [\SFunction{print}, \SFunction{plot}, \SFunction{predict} or
\SFunction{update}] in the same way, regardless of the type of model.
The owner of the software for a particular type of model, on the
other hand, would like to write just that version of each function,
without being responsible for the other versions.

Once stated, this is essentially a prescription for functional OOP: a
class of objects for each kind of model, generic functions for the
computations on the objects and methods for each function for each class.
Where one class of models is an extension of another (analysis of
variance as a subclass of linear models, e.g.), methods can be
inherited when that makes sense.

An implementation of generic functions and methods was introduced as
part of the statistical models project and described in the Appendix
to the white book.
The central mechanism was an explicit method dispatch.
The function \SFunction{print}, for example,
would evaluate the expression:\vspace*{6pt}
\begin{list}{}{\setlength{\leftmargin}{2em}}
\item{}
\begin{alltt}\small
UseMethod("print").
\end{alltt}
\end{list}\vspace*{6pt}
The evaluation of this call would examine the \Cq{class} attribute of
the first formal argument to the function.
If present, this would be a character vector. Eligible methods would
be those matching one of the strings in the class vector; if none
matched, a method matching the string \Cq{default} would be used.
Inheritance was implemented by having more than one string in the
class, with the first string being \dQuote{the} class and the
remainder corresponding to inherited behavior.

\citet{SMS}, in the discussion of classes and methods, noted that \S{}
differed from other OOP
languages because of
its functional programming style.
In fact, this version of functional OOP finessed the resulting
distinction from
encapsulated OOP in two ways.
First, the methods were dispatched according to a single argument, the
first formal
argument of the generic function in principle.
As a result, the methods were unambiguously associated with a single
class, as they would be in encapsulated OOP.
Methods were actually dispatched on either argument to the usual
binary operators, but a number of encapsulated OOP languages
do the same, under the euphemism of operator overloading.

Second, the question of whether methods belonged to a class or a
function was avoided by not having them belong to either.
Methods were assigned as ordinary functions and identified by the
pattern of their name:
``\emph{function}\code{.}\emph{class}''.
In any case, there were no class objects and generic functions were
ordinary functions that invoked \SFunction{UseMethod} to select
and call the appropriate method.
Neither the function nor the class was able to own the methods.

Technically, the method dispatch in this version of OOP was
instance-based, not class-based,
since no rule enforced a consistent set of classes, that is, that
all objects with a given first class string would have identical
following strings for the superclasses.
(\R{} for some time had an S3 class in the base package with a main
class string \Cq{POSIXt},
representing date/times, that could be followed in different objects by one
of two
strings that
in fact represented specializations, i.e., subclasses, of \Cq{POSIXt}.)

The classes and methods implemented for statistical models constituted
a bare-bones version of functional OOP, which is not
to imply that this was a bad idea.
Advantages include a relatively low learning barrier for programming
and a thin implementation layer above the previously existing
language, which in turn means less computational overhead in some
circumstances.
[Interestingly, the encapsulated OOP of \Python{} has a similarly
thin implementation, with classes containing methods but without
defining the properties. A very analogous defense is made for that
implementation, in Section~9 of the \Python{} tutorial,
\citet{Python}, e.g.]

A more formal version of
functional OOP was developed at Bell Labs,
introduced into \S{} in the late 1990s and described in
\citet{SBook98}.
By this time, \S{}-based software was exclusively licensed to the
Insightful Corporation, which later
purchased the rights to the \S{} software, in 2004, and was itself
subsequently purchased by
Tibco.

The new paradigm differed from S3 classes and methods in three main ways:
\begin{longlist}[1.]
\item[1.] Methods could be specified for an arbitrary subset of the
formal arguments, and method dispatch would find the best match to
the classes of the corresponding arguments in a call to the generic function.
\item[2.] Classes were defined explicitly with given properties (the
slots) and optional superclasses for inheriting both properties and
methods.
\item[3.] Generic functions, methods and class definitions were themselves
objects of formally defined classes, giving the paradigm
reflectivity.
\end{longlist}
The new paradigm was part of the version of \S{} described in the 1998
book and generally referred to as S4.
The S4 label is generally applied to this OOP paradigm, whether in \S
{} or
\R{}.
S4 methods never had much chance of replacing S3 methods.
In practice, many S4 generic functions were based on functions that
already dispatched S3 methods.
In this case, the S3 generic function became the default S4 method.

The work on S4 paralleled in time the arrival of \R{} and its
conversion into a broad-based joint project following the initial
publication by \citet{RArticle}.
The implementation of \R{} was designed to provide the functionality
for \S{} described in the blue book and white book, including S3
methods.
Beginning in 2000, an implementation of the S4 version of OOP was added
to \R{}.
The \dQuote{Software for Data Analysis} book,
\citet{R:Chambers:2008}, includes a description of the \R{} version.

Both versions of functional OOP will remain in \R{}.
Many prefer the simplicity of the old form, and in any case the very
large body of existing code will not be discarded, and should not be.
Some important extensions have been made, for example, by registering
the S3 methods from a package.
Major forward-looking projects have typically used the newer version,
for example, the \Language{Bioconductor} project for bioinformatics software,
\citet{Bioconductor}, and the \Language{Rcpp} interface to \Cpp{},
\citet{Rcpp}.
Recent changes, such as making the S3 and S4 versions of inheritance as
compatible as possible, have been aimed at helping the two forms to
coexist productively.

Any programming paradigm with some degree of formality is likely to
have a higher initial learning barrier and require some extra
specification from the programmer.
A comparison of encapsulated OOP programming with \Python{} to that
with \Java{} is an interesting parallel to S3 and S4.
In both examples, the less formal version is likely to be quicker
to learn, while the more formal version provides more information about
the resulting software.
That information in turn can support some forms of validation for the
resulting software, as well as tools to analyze and describe it.
\Python{} and \Java{} being rather different languages in other
respects as well, projects are not too likely to make a choice between
them based
solely on the formality of the object-oriented programming.

With \R{}, a conscious choice is more likely. The arguments for
a more formal approach apply particularly, in my opinion, to projects
with one or more of the characteristics: a
substantial amount of software is likely to be written; the
application has a fairly wide scope in terms of either the data
or the computing methods; or the validity and reliability of the
resulting software is
important.

Nothing prevents good software being written without formal tools in
this case nor of bad software being written with them.
However, there are several potential benefits that can be summarized
in parallel with the main innovations noted above:
\begin{longlist}[1.]
\item[1.] Allowing methods to depend on multiple arguments fits the
functional paradigm
in \R{}, in which the arguments collectively define the
domain of the function.
Many functions in \R{} are naturally applied to different classes of
objects, not necessarily corresponding to the first argument, or only
to one argument.
For example, when binary operators such as arithmetic are defined for
a new class, a clean design of methods for the
operators often needs to distinguish three cases:
the first operand only belonging to the
new class, the second operand only or both operands.
\item[2.] A formal definition for a class allows programmers to
rely on the properties of objects generated from the class. Otherwise,
the nature of the objects can only be
inferred, if at all, from analyzing all the software that creates or
modifies an object of this class.
\item[3.] Having formal definitions for the generic
functions, methods and class definitions themselves supports a growing
set of
tools for installing and using packages that include such functions,
methods or classes.
\end{longlist}
The benefits of a general, reliable form of functional OOP extend to
developments in the language itself.
For example, reference classes were built on the S4 classes and methods,
with no internal changes to the \R{} evaluator required.

\subsection{Reference Classes}
\label{sec:current-status}

Functional OOP remains an active area in \R{}.
In addition, \emph{reference classes},
introduced to \R{} in 2010 in version 2.12.0, provide an
implementation of encapsulated OOP.
Class definitions
include the properties of the class with optional type declarations;
properties may
also be optionally declared read-only.
Class definitions are themselves objects available at runtime.
Methods are programmed as \R{} functions, in which the object itself
is implicitly available, not an explicit argument.
Methods can access or assign properties in the object by name.
These characteristics make the implementation more
\Java{}-like, say, than \Python{}- or \Cpp{}-like.

The programmer defines a reference class in the \R{} style, calling
\SFunction{setRefClass} instead of \SFunction{setClass}.
The call returns a generator for the class and saves the class
definition object as a side effect, as does \SFunction{setClass} for
S4 classes.

As a side comment, while \R{} uses a
model for most of its objects and computations that is fundamentally
different from the object references in encapsulated OOP, a few key features
made the implementation of reference classes in \R{} possible and
even relatively straightforward.
Most importantly, the \R{} data type \SClass{environment} provides a
vehicle for object references and properties.
Environments are universal in \R{} and well supported by programming
tools.
In particular, the active binding mechanism, which allows
access and
assignment operations on objects in environments to be programmed in \R
{}, was valuable in
the implementation.

Reference classes allow the use of encapsulated OOP for objects that
suit that paradigm more naturally than they do functional OOP.
As noted in Section~\ref{sec:from-data-classes}, the essential
distinction between functional and encapsulated OOP is whether an
object is created, once, by a function call or is instead a mutable
object that changes as methods are invoked.

Statistical computing has examples clearly suited to each of these
paradigms.
The linear model returned by \SFunction{lm} is not open to mutation.
Change the numbers in the coefficients or residuals and you no longer
have an object that should belong to that class.
In contrast, a~model simulating a dynamic process such as the
\SExpression{SimplePop} class in Section~\ref{sec:from-data-classes}
exists precisely for the purpose of changing, with its evolution being
the central point of interest.
Other, less directly statistical computations in \R{} also may
correspond to
mutable objects, for example, the frames or other objects in a graphical
interface.

Not every case is clear cut.
Sometimes, essentially the same class structure may be more
appropriate for functional or encapsulated classes depending on the
purpose of the computation.
Data frames are a prime example.
This essential object structure is viewed naturally as functional when
it is part of a functional object related to the data frame.
For example, a fitted model that wanted to be fully reproducible could
return the data frame on which the fitting was based [e.g.,
\SFunction{lm} includes the model frame it constructs].
Such a data frame is clearly functional; again, change it and you
invalidate the model.
On the other hand, a data frame to be used in data cleaning and
editing is an object that needs to be mutable.

Having both paradigms in a single language is unusual.
Some functional-style languages have implemented functional
OOP, notably \Language{Dylan}, interesting for its parallels with
OOP in \R{}---see \citet{Dylan}, particularly the discussion of method
dispatch.
Other languages with a functional structure have nevertheless added
what is essentially
encapsulated OOP, for example, \citet{Scala} for the case of
\Language{Scala}.

We hope that providing both paradigms in \R{} encourages software
design that is natural for
the application.
It does at the same time pose some subtleties.
Reference classes and reference class objects are somewhat
abnormal in \R{}.
One needs to understand the distinctions from standard \R{} objects.

The key is the local reference mechanism noted in Section~\ref{sec:in-R}.
The \R{} evaluator enforces local reference by duplicating an object
when a computation might alter a nonlocal reference.
Certain object types are exceptions that are not duplicated.
The important exception is type \Cq{environment}. Reference classes
are implemented by extending this type.
Encapsulated OOP in \R{} uses no special form of the function
call.
Method invocation is just a call to the
\SOperator{\$}
operator, for which reference classes have an S4 method.
Reference semantics are obtained by one basic fact: environments are never
duplicated automatically.
The S4 class mechanism in \R{} nevertheless allows one to subclass the
\Cq{environment} type in order to define reference class behavior.

The objects in the fields of a reference class object can be ordinary
\R{} objects.
They behave just as usual and when used in function calls will
have regular local reference behavior in that call.
It is only when fields in the reference object itself are replaced
that the encapsulated OOP is relevant.

Reference class objects are also good candidates for interfaces to
other languages that implement the same OOP paradigm, such as \Java{},
\Cpp{} or \Python{}.
The \R{} object could be a proxy for an object in the other language
with methods invoked in \R{} but executed on the original object.
The \SPackage{Rcpp} interface to \Cpp{}, \citet{Rcpp}, has a
mechanism for extending
\Cpp{} classes in this way.
\Cpp{} classes can only be
inferred from the source, meaning that either the programmer must
supply the interface information (as in the current implementation) or
some processing of the source must be applied (currently used to
export functions from \Cpp{} but not classes).
\Java{} classes are accessible as objects, via \dQuote{reflectance} in
\Java{} terminology, so that in principle proxy classes in \R{} should
be possible.
The \SPackage{rJavax} package by \citet{rJavax} has an initial
implementation.
For \Python{}, methods are available from the objects but properties
are not formally defined.
At the time of writing, basic interfaces to \Python{} exist, for
example, \citet{rJython}, which could be extended to support class
interfaces, with methods but not properties inferred from the
\Python{} class objects.

Further work on these and other inter-system interfaces would be a
valuable contribution to the user community.

\section{Summary}
\label{sec:summary}

\R{} plays a major role in the communication and dissemination of new
techniques for statistics and for results of statistical research more
generally.
In particular, the many
packages written in \R{} or using \R{} as a base for interfacing to
other software
constitute an essential, rapidly growing resource.
Therefore, the quality of such software and the ability of programmers
to create and extend it are important.

The current \R{} language and its supporting functionality are the
result of many years of evolution, from early programming libraries
through the \S{} language to \R{}, which itself has evolved and
accumulated a variety of programming techniques.
This evolution has been much influenced by the functional and
object-oriented programming paradigms.
New versions have
continued to include supporting software and programming tools found
useful at earlier stages along with improved capabilities.

The programming paradigms become especially relevant when the
applications are complex or the quality of the resulting software is
important.
In particular, the versions of object-oriented programming in \R{} can
assist in dealing with complexity of the underlying data.
As noted, \R{} implements OOP in two forms, functional and
encapsulated.
These are complementary, with one or the other suitable for particular
applications.
The latter is essentially the form of OOP used in most other
languages, but the former is distinctly different.
Considerable confusion has arisen in discussions of OOP in \R{} from
not noting that distinction, which the present paper has tried to clarify.

More generally, understanding the role of object-oriented
and functional programming in \R{} may assist future contributing
programmers in using related programming tools.
The continuing rapid growth of \R{}-based software and the expanding,
challenging range of techniques it has to support make effective
programming an important goal for the statistical community.

The importance of object-oriented programming is likely to increase as
statistical software takes on new and challenging applications.
In particular, the need to deal with increasingly large objects and
distributed sources of data will bring in specialized classes of data
and will need powerful computing tools.
One important direction has been to transform selected software in \R{},
particularly to speed up large-scale computations; see, for example,
the companion paper \citet{DTLpaper}.
Complementary to this is to interface to other languages
and software when these provide better performance on \dQuote{big
data} and other computationally demanding applications.
In particular,
interfaces that match with object-oriented treatments for specialized
forms of data can exploit the OOP facilities in \R{}.
The interface to \Cpp{}, \citet{Rcpp}, is an example.
Further development of such interfaces will be of much benefit.

Functional programming is perhaps not such an obviously hot topic at
the moment.
However, the underlying philosophy that our software should be in the
form of reliable, defensible units is very much part of \R{}.
Situations where the validity of statistical computations needs to be
defended are likely to increase, given the growing need for
statistical treatment of complex problems for science and society.

\section*{Acknowledgments}
\label{sec:acknowledgements}

Thanks to the Associate Editor and the referees for some helpful
comments on presentation and content.
Thanks especially to Vincent Carey for organizing and encouraging the
set of talks and papers of which this is part.

%



\end{document}